\documentclass[12pt,preprint]{aastex}

\shorttitle{The Cooling of WZ Sge}
\shortauthors{Godon et al.} 

\begin{document}

\title{Modeling the Heating and Cooling of WZ Sagittae
following the July 2001 Outburst.}

\author{Patrick Godon$^1$, Edward M. Sion, \& Fuhua Cheng}
\affil{Department of Astronomy \& Astrophysics, Villanova
University, Villanova, PA 19085;
patrick.godon@villanova.edu emsion@ast.villanova.edu}

\author{Boris T. G\"ansicke} 
\affil{Department of Physics \& Astronomy, University of
Southampton, Highfield, Southampton SO17 1BJ; 
btg@astro.soton.ac.uk}  

\author{Steve Howell} 
\affil{Institute of Geophysics \& Planetary Physics, 
1432 Geology Building, University of California, 
Riverside, CA 92521; 
steve.howell@ucr.edu}  

\author{Christian Knigge} 
\affil{Department of Physics \& Astronomy, University of
Southampton, Highfield, Southampton SO17 1BJ;
christian@astro.soton.ac.uk}  

\author{Warren M. Sparks}
\affil{XNH, MS F664, Los Alamos National Laboratory, Los Alamos,
NM 87545; wms@lanl.gov} 

\author{Sumner Starrfield} 
\affil{Department of Physics \& Astronomy, Arizona State University,
Box 871504, Tempe, AZ 85287-1504;
sumner.starrfield@asu.edu}  

\altaffiltext{2}
{Visiting at the Space Telescope Science Institute, Baltimore, MD 21218, USA}

\clearpage 

\begin{abstract}

FUSE and HST/STIS spectra of the dwarf nova WZ Sge, obtained
during and following the early superoutburst of July 2001
over a time span of 20 months, monitor changes
in the components of the system during its different phases.
The synthetic spectral fits to the data
indicate a cooling in response to the outburst of about 12,000K,
from $\approx 28,000K$ down to $\approx 16,000K$. 
The cooling time scale $\tau$ (of the white dwarf temperature excess) 
is of the order of $\approx 100 $
days in the early phase of the cooling period, 
and increases to $\tau \approx 850$ days   
toward the end of the second year following the outburst. 

In the present work, we numerically model the accretional heating
and subsequent cooling of the accreting white dwarf in WZ Sge.
The best compressional heating model fit is obtained for a
$1.2 M_{\odot}$ white dwarf accreting at a rate of $9 \times 10^{-9}
M_{\odot}$yr$^{-1}$ for 52 days. 
However, if one assumes a lower mass accretion rate or a lower
white dwarf mass, then compressional heating 
alone cannot account for the observed temperature decline, and 
other sources of heating have to be included to increase
the temperature of the model to the observed value. We quantitatively
check the effect of boundary layer irradiation as such an additional
source.  

\end{abstract}


\section{Introduction}

WZ Sge is a well-studied dwarf nova
(DN, a sub-class of Cataclysmic Variables - CVs) 
with extreme properties. It has the largest
outburst amplitude, shortest orbital period, longest outburst
recurrence time, lowest mass Roche-lobe filling secondary, 
and lowest
accretion rate of any class of DNe \citep{how99,how02}.
In addition it is the
brightest DN and arguably the closest CV,
with a distance of only $\approx 43$ pc (Thorstensen 2001, private
communication). WZ Sge was observed to go into outburst in 1913,
in 1946 and in 1978 and it was consequently assumed to have an outburst
period of about 33 years.
The 23 July 2001 outburst, first reported by T. Ohshima (see \citet{ish01}),
was therefore 10 years earlier than expected.

Following the nomenclature of \citet{pat02},  
the optical light curve (Figure 1) 
of the system during the July 2001 outburst  
exhibits a "plateau" phase (this phase is really 
a slow decline phase), which lasted for about 24 days. 
During this period the system brightness underwent a steady decline
with a rate of about 0.1 mag/day, falling from $M_v \approx 8.2$ to 
$M_v \approx 10.7$ in about 24 days, a sign that the accretion was 
taking place at a slowly decreasing rate. 
It was then followed by a sharp drop (itself lasting a few days)
of the visual magnitude from $m_v \approx 11$ down to $m_v \approx 13$,  
where it stayed for about 3 days: the "dip". 
During the dip, accretion had either stopped completely or 
dropped considerably.  
Between day 29 and day 52 of the outburst, the system then
underwent 12 successive ``echo'' outbursts: the "rebrightening" phase.
On the 53rd day, the system started to cool without
any other noticeable outburst event: the "cooling" phase
(in Figure 1 the light curve is shown for $t<75$ days only, since
for $t> 53$ days $M_v$ is a monotonously decreasing function of
time, as the star cools). \\  

The 2001 outburst light curve of WZ Sge is remarkably similar to 
its 1978 outburst light curve with practically the same
initial decline rate of 0.1mag/day 
\citep{bai79}. However, in 1978 the plateau and
rebrightening phases lasted about 30 days each against 25 days each
in the 2001 outburst
(i.e. the outburst in 1978 lasted about 10 days longer than in 2001). 
This may be related to the fact that in  
2001 WZ Sge erupted after only 23 years of quiescence, while in
1978 it erupted after 33 years of quiescence. The
2001 outburst was not as strong as in 1978.  
From the IUE archive we found that the continuum level of the spectrum 
has a flux density $F_{\lambda}$ (estimated around $\lambda = 1,700$ \AA\ )
of about $ F_{\lambda} \approx 2 \times 10^{-11}$ ergs s$^{-1}$ cm$^{-2}$ 
\AA $^{-1}$ during the early phase of the 1978 outburst, or about
3 times larger than in the 2001 outburst 
at the same epoch and wavelength  
$F_{\lambda}\approx 7 \times 10^{-12}$ ergs s$^{-1}$ cm$^{-2}$ \AA $^{-1}$. 
During the cooling phase, the flux density $F_{\lambda}$ was also larger 
in 1978 than in 2001 at the same epoch \citep{sle99}.  
The 1946 eruption,
however, was much different, with no apparent rebrightening phase, 
while the plateau phase had small amplitude variations 
during the whole duration of the outburst. \\   

The main purpose of the present work is to 
try to account for the observed sequence
of temperatures using a numerical model 
for the accretional heating and subsequent cooling of the 
accreting white dwarf. 
In the next section we address the issue of the accuracy of
the temperature determination during the
outburst and cooling periods,   
together with an overview of the estimates of the
WD temperature and accretion rate. 
We present the code we use
to carry out the simulations of the heating and
cooling of the WD in section 3.  
The results are presented in section 4. 
In the last section we discuss the possible origin of the observed 
high temperatures and slow cooling of the white dwarf. 

\section{The Accretion Rate and White Dwarf Temperature Determination}

In the present work we consider the observations of WZ Sge carried 
out in the
Far Ultraviolet for which the temperature of the accreting white
dwarf and/or the mass accretion rate of the system were assessed.
These observations monitor changes in the FUV component of the 
system during its different phases, over a time span of 20 months. 
and reveal a cooling of the WD in response to the outburst of 
about 12,000K.
Namely, we consider 4 FUSE and 11 HST/STIS observations as
follows:
one FUSE observation obtained during the plateau phase,
one HST observation obtained during the dip, two HST and one FUSE
observations obtained during the rebrightening phase and 2 FUSE
with 8 HST observations obtained during the cooling phase.
A total of 15 observations together with their references
are recapitulated in Table 1.

In order to estimate the temperature  
one generates a grid of theoretical spectra for different 
values of the surface temperature, gravity and composition,
using the synthetic spectra generator codes TLUSTY and SYNSPEC 
\citep{hub88,hub94,hub95}.  
One then masks regions of the observed spectrum that are not 
characteristic
of a WD atmosphere (such are emission lines for example),  
and a $\chi^2$ fitting technique is used to find the best
fit model. The mass accretion rate is estimated in a similar
manner using different options in the synthetic spectra
generator codes TLUSTY (the older version is known as TLDISK) 
and SYSNPEC.  

{\bf{Mass Accretion Rates:}} 
the following mass accretion rates were estimated in the
references mentioned in Table 1 and should be considered
only as an order of estimate. 
From Table 1, 
we see that the mass accretion rate estimate during the
plateau phase is rather modest, of the order of
$1-3 \times 10^{-9}M_{\odot}/year$ 
\citep{lon03a} on day 7 into outburst (Table 1). 
During the dip, the white dwarf 
might have a temperature of about $25,000$K 
\citep{kni02} on day 27. 
This is more than $10,000K$ above its
quiescence temperature which is $\approx 14,500K$. 
During the rebrightening
phase, the mass accretion rate seems to be somewhat smaller than during the
plateau and only peaks at $1\times 10^{-9} M_{\odot}$yr$^{-1}$ 
\citep{lon03a} on day 46. 
We put these
values into the table, though we feel that they should be considered only
as rough estimates rather than actual values and they   
reflect our inability to assess $\dot{M}$ (and $T_{wd}$) during this epoch. 
This is because during the different phases of the outburst,  
additional components (such as the
accretion disk and other obscuring material ejected during outburst)  
contaminate and veil  \citep{lon03b}
the spectrum of the white dwarf. 
However, during the cooling phase, one expects to 'see'
mainly the white dwarf with little or no contribution from the disk
and/or other (possibly masking) additional components. 
During this epoch the white dwarf is
exposed and its temperature decreases.

Because of that, we decide to use for our modeling the values of
the white dwarf temperature obtained during the cooling phase
only. However, discrepancies of up to 5,000K exists 
between the temperature estimates
of \citet{sio03} and \citet{lon03b},
for the observation carried out
on day 50, when the white dwarf was revealed, but apparently also
partially masked or/and possibly with an accretion disk component. 
The rest of the observations in the cooling phase 
are consisten within 2,000K. 

{\bf{Temperatures:}} 
we elaborate here a little more on the temperature, 
using the three different approaches (denoted a, b and c
in Table 2). 

These were all observations using HST/STIS.  
In (a) we mask the N V region of the spectrum when needed (together with
less affected regions varying from spectrum to spectrum ) and use the 
latest version of the 
synthetic spectra generator codes TLUSTY and SYNSPEC 
\citep{hub88,hub94,hub95}; in
(b) the spectral fitting technique uses the same
masking regions for all the spectra together with the new version
of the code; and (c) is taken from \citet{lon03b},
which uses a third and different masking technique together with
an earlier version of the code. In Table 2, we also list the flux
integrated over the wavelength over the entire spectral range of
STIS, to the power 1/4, for all the epochs, which is proportional to
the effective temperature (though each integrated flux can be relatively
over- or under- estimated).  

In Figure 2 we draw the temperatures listed in Table 2 as follows: 
$T_a$ is represented by stars; $T_b$ is represented by squares;
$T_c$ is represented by plus signs; and the temperature estimated
through the flux is represented by circles, where it has been arbitrarily
scaled so as to fit the last data point for $T_b$. The two triangles
represent the FUSE data points. The September 2001 datum
has temperature estimates
ranging from $\approx 27,000$K up to $\approx 32,000$K. For that particular
epoch the spectrum was taken between the last two echo outbursts and
obviously the white dwarf was not the only component and in addition
might have been partially masked. For that reason we decide not to   
take into consideration the data points that were not 
known accurately, or which were 
taken during periods when the white dwarf might still have been veiled. 
In our numerical modeling we will try to model the cooling temperature 
as listed in Table 2, with some reservations about the September 2001
entry.  

The cooling of the white dwarf is initially pretty rapid and then 
slows down significantly. If we assume an exponential
cooling law $T = T_{inf}+\Delta T \times e^{-t/\tau}$ 
(with $T_{inf} = 14,500$K), equivalent to  
$\Delta T(t) = \Delta T \times e^{-t/\tau}$, where
$\Delta T(t)$ is the excess temperature of the WD at time $t$, then
we find that $\tau$ is not constant, i.e. the cooling is not
exponential.
In Tables 1 and 2, $t$ is counted from the begining of the cooling period,
day 52 into outburst, however, it is usually agreed to count the
cooling time from the beginning of the cooling phase. Therefore,
we set here $t \rightarrow t-52$ days, and   
we denote the cooling time 
at time $t= n~days$ by $\tau^n$. We find 
$\tau^{13} \approx (70~^+_- 10 )$ days, 
$\tau^{50} \approx (190~^+_- 30 )$ days, 
$\tau^{300} \approx (325~^+_- 40 )$ days, 
and $\tau^{550} \approx (850~^+_- 280 )$ days. 

It is interesting to note that 
the observed cooling following the 2001 July outburst is faster than
the cooling observed after the 1978 outburst. 
\citet{sle99} found a WD temperature of 20,500K, 220 days after the outburst
against $\approx$ 19,000K at the same epoch for the 2001 outburst 
(e.g. observing epoch of April 2002, day 266). 
And $\approx$ 500 days after the 1978 outburst the temperature was still 
around 17,500K while the present observations seem to indicate a 
lower temperature ($\approx$ 17,000K) 466 days after the 2001 outburst. 

\section{Numerical Modeling}

In order to model the accretional heating and subsequent cooling
of the WD in WZ Sge, 
we use a 1D quasi-static evolutionary code without hydrodynamics
(quasi-static assumption).
It is an updated version of the quasi-static stellar
evolution code of \citet{sio95}.
The code includes time variable accretion, OPAL opacities and 
boundary layer irradiation which indirectly accounts for the 
stellar rotational velocity (see Eq.1 below).
Stellar rotation, however, is not included anywhere else in the code. 
In this code, the white dwarf is computed all the way down to the
core, in the region well below the nuclear burning region.
Initial models are constructed by the fitting point method and the
resulting initial model down to a depth where $\rho =10^6 g~$cm$^{-3}$ 
is stored
as input for the evolution code.
For a given white dwarf mass and an initial effective surface temperature
($T_{eff}$), we chose  $R_{wd}$ to be the theoretical equilibrium radius.
The models built in this manner are initially
in equilibrium, and if no matter is accreted onto the WD surface, the
WD parameters do not undergo any change on the time scale studied here.
All other details of the code can be found in \citet{sio95}
and references therein.

Numerical simulations
are carried out by switching on accretion
for the duration of a dwarf nova outburst interval and then shutting it
off to follow the cooling of the white dwarf.
In this way the effects of compressional heating
and boundary layer irradiation can be assessed quantitatively.
The matter is assumed to accrete 'softly' with the same
entropy as the white dwarf outer layers. 
From theoretical considerations, one expects the accreting
matter, as it transits through the boundary layer region, 
to increase its temperature (since the boundary layer temperature 
is high: $T_{BL} >> T_{*}$). In addition in the boundary layer
region, the radial velocity is large and the accreting matter can
also heat up by shock. For these reasons the 'soft' accretion
assumption is actually not justified, since this advected energy
is non-negligible. However, in our treatment of the  
boundary layer irradiation, we take into account part of this energy
by assuming that a fraction of the energy liberated in the boundary layer
($L_{BL}$) is absorbed by the outer layer of the star: it is included as 
a source in the energy equation in the outer layer of the star. 

The treatment of the boundary layer irradiation is done as follows. 
The energy liberated in the boundary layer 
is given by \citep{klu87}:
\begin{equation}
L_{BL}=\frac{1}{2} L_{acc} \left( 1 -
\frac{\Omega_{*}}{\Omega_K(R_{*})} \right) ^2,
\end{equation}
where
\[
L_{acc}=\frac{G M_{*} \dot{M}}{R_{*}}
\]
is the total accretion energy, $G$ is the gravitational constant,
$M_{*}$ is the mass of the star, $R_{*}$ is the radius of the star,
$\dot{M}$ is the mass accretion rate, $\Omega_{*}$ is the angular
rotation
rate of the star and $\Omega_K(R_{*})$ is the Keplerian angular
velocity at one stellar radius. 
Equation (1) can be used as long as the disk is
geometrically thin and optically thick and extends to
the stellar surface. 
In the present case one expects the stellar rotation
rate to be the rotation rate of the WD: $\Omega_{*}= \Omega_{wd}$.

We assumed that only a fraction of
the boundary layer luminosity is irradiating the star,
namely:
\[
L_{irr} = \alpha \frac{L_{BL}}{2}.
\]
A value $\alpha=1$ means that half of the BL luminosity is
lost into space while the other half is absorbed by the star. 
Assuming a value $\alpha=0.5$ means that only 25\% of the BL luminosity
is absorbed by the star. 
Here, we choose $\alpha=0.5$, which is the value used
in the work of \citet{sha87}, and about half the estimated
value of \citet{reg89} who used $\alpha \approx 0.2 L_{acc}$.
We then assume different
rotation rates
\[
\eta = \frac{\Omega_{*}}{\Omega_K(R_{*})}.
\]
For a non rotating star $\eta = 0$ and the
BL luminosity is exactly half the accretion energy $L_{BL}=L_{acc}/2$,
while for a star rotating near break-up $\eta=1$ and $L_{BL}=0$.  
In this work we took values  
ranging from $\eta = 1$ (for no boundary layer irradiation, in order to
check only the effect of the compressional heating) down to
$\eta = 0.05$ (when the star is slowly rotating and boundary
layer irradiation is the main source of heating). Clearly a smaller value of
$\alpha$ will require a smaller value of $\eta$ in order to keep the
same amount of BL irradiation in a specific model, and vice versa.  
However, the compressional heating results obtained in this work are not 
at all affected by the value of $\alpha$ used in the simulations. 

In the simulations it is assumed that the accretion and 
heating of the white dwarf 
is uniform rather than being restricted to the equatorial region. 
In addition, the transfer of angular momentum (by shear mixing) 
into the white dwarf is neglected.

\section{Results}

As previously remarked in section 2, 
during the outburst, as accretion takes place at a
high rate, 
the star's photospheric emission is overwhelmed by the emission of 
the hot components (mainly the inner disk),
which makes it difficult to assess the exact temperature of the star
and its rotation rate $\Omega_{wd}$.
However, on day 53 the system is found in a low state
and starts to cool. During that time the accretion rate has probably
dropped to its quiescence level.

Numerical simulations \citep{god02} have shown that the 
temperature increase, due to BL irradiation, is sustained
only during accretion, and when the accretion is turned off, the star
rapidly radiates away the BL 
energy absorbed in its outermost layer. However,
the temperature increase due to compressional heating takes place deeper
in the outer layers of the star and it takes many days (months) for
the star to cool. 
Therefore, we assume that in the cooling phase the observed 
elevated temperature of the star is due to the compressional heating it
has endured during the outburst phase alone. 

Since BL irradiation and compressional heating take place at different
depths in the outer layers of the star, and on different time scales, 
we model their effect on the temperature of the star in separate model 
runs. 

First we model the compressional heating of the star alone ocuring
during the outburst phase,
therefore setting $\eta =\Omega_{wd}/\Omega_K =1 $ in the simulations. 
The exact mass accretion
rate of the system is not known and it is changing with time during the
outburst. 
From the optical light curve in Figure 1, it seems very likely 
that initially the
mass accretion rate is very high at the onset of the outburst and
decreases steadily during the plateau phase.
At present, our code only simulates a constant accretion
rate, but at any time the accretion can be turned off to model the cooling.   
Consequently, we model the outburst phase (plateau + rebrightening) by 
accreting at a constant rate for 52 days, after which accretion is
turned off and the WD cools.  
From Table 1, we expect the mass accretion
rate during the outburst phase to be of the order of (say) $3\times 10^{-9}
M_{\odot}$yr$^{-1}$ and the temperature of the star to 
increase from about $15,000K$ up to $28,000K$ (or more, see
Table 2) around day 53. 
We run models with different white dwarf mass, namely $M_{wd}
= 0.7 M_{\odot}$, $1.0 M_{\odot}$,  $1.2 M_{\odot}$ and vary 
the mass accretion rate in the range 
$ 10^{-9}M_{\odot}$yr$^{-1} < \dot{M} < 10^{-8}M_{\odot}$yr$^{-1}$. 
We chose the initial WD temperature to be around $13,500-15,000K$.
Ideally we are looking for a model that reaches a temperature of
$28,000K$ (or more) after 52 days of steady accretion, with $\dot{M} \approx 
3 \times 10^{-9} M_{\odot}$yr$^{-1}$, and with a temperature curve 
as close as possible to the observations of Tables 1 \& 2. 
In Table 3 we list models (1-4) obtained for different values
of $M_{wd}$ and $\dot{M}$. 
From these models, model 3 with $M_{wd}=1.2 \times M_{\odot}$
and $\dot{M} = 9 \times 10^{-9} M_{\odot}$/year,  
yields the best fit. Both models 1 \& 2, with a lower white dwarf mass 
($0.7 M_{\odot}$ and 1 $M_{\odot}$ respectively),
and model 4, with a lower mass accretion rate ($5 \times 10^{-9}
M_{\odot}$/year), yield a lower temperature and a sharper
initial temperature drop than the observations. 
In figure 2 we draw the temperature increase of the WD due to 
compressional heating alone, together with the observed values, as a function 
of time for model 3. Models 4 \& 2 are drawn in Figures 3 \& 4 respectively.

Clearly, from Figures 3 \& 4, we see that 
compressional heating alone cannot account for the observed temperatures
if the mass $M_{wd} < 1.2 \times M_{\odot} $ and/or 
the mass accretion rate $\dot{M} < 10^{-8} M_{\odot}$/year, 
because it produces lower temperatures than observed.  
Since both a $1.2 M_{\odot}$ mass and a $10^{-8} M_{\odot}$/year
accretion rate are larger than expected, we decided to check
the effect of boundary layer irradiation on the temperature
of the white dwarf for a model with $M=1.0 M_{\odot}$ in addition
to the compressional heating. Namely, we decide to check how 
boundary layer irradiation increases the temperature of
the model in addition to compressional heating.  
Specifically, we chose model 2 arbitrarily around t=175 day.
There it has a temperature
of only 18,000K (due to compressional heating, Figure 4) while the 
observations  indicate that
the temperature should be at least 2,000K higher at that epoch.  
We therefore run a 1.0 $M_{\odot}$ model (model 5, same mass as model 2)
with an initial temperature of 18,000K (to simulate day 175 of model 2) 
and a quiescent mass accretion rate 
($\dot{M} \approx 10^{-12}-10^{-11} M_{\odot}$/year), 
and we include boundary layer irradiation assuming a rotation rate
$\Omega_{wd}=0.2 \Omega_K$ (corresponding to the observed rotational 
velocity of 1200km/s). Namely, we wish to check if the additional heating
of the boundary layer irradiation of the accretion disk during the cooling 
phase (quiescent $\dot{M}$) can increase the temperature of
the model by 2,000K, and for what value of $\dot{M}$ (here we model
the combine effect of compressional heating and boundary layer
irradiation separately, model 2 and model 5 respectively).  

We find, from model 5 in Table 3, that one can increase the temperature
of model 2 at t=175 days by 2,000K by means of   
boundary layer irradiation if the mass accretion rate
is at least $\dot{M} \approx 2 \times 10^{-11} M_{\odot}$/year.

\section{Discussion and Conclusion}

In the present numerical exploration we first assumed that all
the accretional heating of the white dwarf was due to compressional
heating, and we found that the mass of the accreting 
white dwarf needed in the 
simulations of the WZ Sge superoutburst to fit the observations
was rather large: $1.2 M_{\odot}$.
The average mass accretion rate of the
outburst model, $9 \times 10^{-9} M_{\odot}$yr$^{-1}$, 
assessed from the compressional 
heating simulations, was itself larger by a factor of $\approx 5$, than the 
average value determined from the spectral fits to the observations
which was $\approx 1.7 \times 10^{-9} M_{\odot}$yr$^{-1}$ [25 days of 
accretion at a rate $\dot{M}= 3 \times 10^{-9} M_{\odot}$yr$^{-1}$, 
followed by 25 days of intermittent accretion
at an average rate of  $\approx 5 \times 10^{-10} M_{\odot}$yr$^{-1}$].  

When we assumed a smaller WD mass of only one solar mass,
we found that even with a high mass accretion rate of 
$10^{-8} M_{\odot}$/year the
model could not account for the observed temperature. 
For that model, we simulated a second source of
heating (in addition to compressional heating): 
boundary layer irradiation at a quiescence 
mass accretion rate, corresponding to the mass accretion
rate during the cooling period. 
The mass accretion rate we obtained 
($2 \times 10^{-11} M_{\odot}$/yr; Table 3) to fit the
observations   
is in agreement with a recent analysis of an X-ray observation
of WZ Sge. \citet{has02} presented results of a uniform analysis
of all the {\it{ASCA}} X-ray observations of non-magnetic CVs. 
For WZ Sge they estimated an X-ray luminosity  of about 
$L_X \approx 2.7 \times 10^{30}$ erg s$^{-1}$ in the range 0.5-10
keV, assuming a distance of 69 pc. Rescaling this value to the
better estimate of 43 pc (and also to be consistent with the distance
assumed in this work) leads to $L_X \approx 7 \times 10^{30}$
erg s$^{-1}$ for the X-ray Luminosity. \citet{has02} 
found that optically thin boundary 
layer models \citep{pop99} provide the best description of the data. 
Since the disk is radiating in both +z and -z directions, the X-ray
luminosity is at least half the boundary layer luminosity,
namely: $L_X = L_{BL}/2$, and if the star does partially mask
the inner disk and boundary layer (since $i=78^o$), then one
has $L_X < L_{BL}/2$. For this reason we assume 
$L_{BL} \approx 3 \times L_X \approx 2 \times 10^{31}$ erg s$^{-1}$, 
and using
\citet{klu87}'s relation for $L_{BL}$, 
an angular velocity of $1200$km s$^{-1}$ and 
a $1.0$ solar mass, we find the quiescence mass accretion rate
from the X-ray observation to be 
$\dot{M} = 6 \times 10^{-12} M_{\odot}$/yr, about 3 times smaller than
our estimates.  

Another possibility, that we are unable to assess quantitatively,
is the slow release of rotational kinetic energy from the outer layer of
the star \citep{spa93}. If the outer layer of the star (equatorial belt) has been 
spun up during the outburst, then rotational kinetic energy can be release
during the cooling phase as this fast rotating layer spins down. This
effect could account for an additional source of heating of the white
dwarf ocuring during the cooling phase.  

In this work we have shown that compressional heating alone can account
for the cooling curve of WZ Sge following the July 2001 superoutburst
only if the WD in WZ Sge is massive ($M_{wd}=1.2 M_{\odot}$) and the outburst
accretion rate is large ($\approx 10^{-8} M_{\odot}$yr$^{-1}$). 
If this is not the case, then compressional heating alone is not 
enough to account for the observed decrease in the WD temperature, and 
we suggest boundary layer irradiation from a quiescent accretion
disk and the slow release of rotational kinetic energy from a
fast rotating accretion belt as possible additional sources of
heating of the white dwarf to account for the observed temperatures.

\acknowledgments

We are thankful to Knox Long for providing us with temperature estimates
from his work before publication. 
We are grateful to the VSNET collaboration for providing us with the data shown
on Figure 1. BTG acknowledges support from a PPARC Advanced Fellowship.  
P. Godon is particularly thankful to C. Hasenkopf for providing us with
a copy of her paper on the X-ray analysis and to Thierry Lanz for kindly
providing us with the latest version of the Tlusty and Synspec codes. 
S. Starrfield acknowledges partial support from NSF and NASA Grants to ASU.  

{} 

\clearpage

\begin{deluxetable}{ccrccccc}
\tablewidth{0pc}
\tablecaption{
FUSE \& STIS observations of WZ Sge
}
\tablehead{
obs. & date      & day & instrument & $T_{wd}$   & $\dot{M}$        & phase  &
Reference$^1$ \\
number &           &     &            & $<1,000K>$ & $<M_{\odot}/yr>$ &        &
\\
}
\startdata
1 & 30 Jul 01 &   7 & FUSE       &   -        &  1-3e-09         & plateau
& [1]   \\
2 & 19 Aug 01 &  27 & STIS       &  25.0      &     -            & dip
& [2]   \\
3 & 22 Aug 01 &  30 & STIS       &   -        &  5e-10           &
rebrightening$^2$ & [2]   \\
4 & 07 Sep 01 &  46 & FUSE       &  42.0      &     -            &
rebrightening$^2$ & [1]   \\
4 & 07 Sep 01 &  46 & FUSE       &   -        &  1e-09           &
rebrightening$^2$ & [1]   \\
5 & 11 Sep 01 &  50 & STIS       &  Table 2   &  3e-10           &
rebrightening$^2$ & [3]   \\
6 & 29 Sep 01 &  68 & FUSE       &  25.0      &    -             & cooling
& [1]   \\
7 & 10 Oct 01 &  79 & STIS       &  Table 2   &    -             & cooling
& [3]   \\
8 & 07 Nov 01 & 107 & FUSE       &  23.0      &    -             & cooling
& [1]   \\
9 & 10 Nov 01 & 110 & STIS       &  Table 2   &    -             & cooling
& [3]   \\
10 & 11 Dec 01 & 141 & STIS       &  Table 2   &    -             & cooling
& [3]   \\
11 & 15 Apr 02 & 266 & STIS       &  Table 2   &    -             & cooling
& [3]   \\
12 & 05 Jun 02 & 317 & STIS       &  Table 2   &    -             & cooling
& [3]   \\
13 & 27 Aug 02 & 400 & STIS       &  Table 2   &    -             & cooling
& [3]   \\
14 & 01 Nov 02 & 466 & STIS       &  Table 2   &    -             & cooling 
& [3]   \\
15 & 23 Mar 03 & 608 & STIS       &  Table 2   &    -             & cooling 
& [3]   \\
\enddata
\tablenotetext{1}
{References: [1] Long et al. 2003a; [2] Knigge et al. 2002; [3] 
see Table 2.}
\tablenotetext{2}
{The observations on day 30 and day 46 coincide with the first and 10th echo 
outburst of the rebrightening phase respectively.  The observation made 
on day 50 was made during a relatively 'low' state between the 11th and 
the 12th outburst in the rebrightening phase.  The observation on day 
46 was modeled twice: once with a white dwarf atmosphere only and once 
with an accretion disk only. The observation on day 50 was modeled
once with a white dwarf atmosphere only and once as a combination of an 
accretion disk and a white dwarf atmosphere.}
\end{deluxetable}

\clearpage

\begin{deluxetable}{ccrcccc}
\tablewidth{0pc}
\tablecaption{
STIS temperature estimates of WZ Sge
}
\tablehead{
obs.   & date & day & $T_{a}^1$ & $T_{b}^1$ & $T_{c}^1$ & (Flux)$^{1/4}$ \\
number &      &          &$<1000K>$&$<1000K>$&$<1000K>$& \\ }
\startdata
 5 & 11 Sep 01 &  50 &  31.9 & 27.0 & 28.2 &  3.8x$10^{-3}$ \\ 
 7 & 10 Oct 01 &  79 &  25.2 & 23.6 & 23.4 &  3.3x$10^{-3}$ \\ 
 9 & 10 Nov 01 & 110 &  23.7 & 22.4 & 22.1 &  3.1x$10^{-3}$ \\ 
10 & 11 Dec 01 & 141 &  22.6 & 21.6 & 20.7 &  2.9x$10^{-3}$ \\ 
11 & 15 Apr 02 & 266 &  19.5 & 18.8 & 18.1 &  2.5x$10^{-3}$ \\ 
12 & 05 Jun 02 & 317 &  18.8 & 18.0 & 17.4 &  2.5x$10^{-3}$ \\ 
13 & 27 Aug 02 & 400 &  17.8 & 17.4 & 16.7 &  2.4x$10^{-3}$ \\ 
14 & 01 Nov 02 & 466 &  17.5 & 17.0 & 16.3 &  2.3x$10^{-3}$ \\ 
15 & 23 Mar 03 & 608 &  17.2 & 16.6 & 15.9 &  2.2x$10^{-3}$ \\ 
\enddata
\tablenotetext{1}
{$T_a$ and $T_b$ were estimated in this work (see text) while
$T_c$ is from Long et al. 2003b.} 
\end{deluxetable}

\clearpage

\begin{deluxetable}{cccccccc}
\tablewidth{0pc}
\tablecaption{
Accretional heating models of WZ Sge
}
\tablehead{
Model&$M_{wd}$&$Log
R_{wd}$&$T_{wd}^{i(1)}$&$\dot{M}$&$T_{wd}^{max(2)}$&$\Omega_{wd}$&length$^{(3)}$\\
number &$<M_{\odot}>$&$<cm>$ & $<1,000K>$ & $<M_{\odot}/yr>$ &$<1,000K>$
&$<\Omega_K>$&$<days>$\\
}
\startdata
1 & 0.7 & 8.95    & 15.0   &  1e-08     & 21.0 & 1.0  & 52    \\
2 & 1.0 & 8.78    & 15.0   &  1e-08     & 25.0 & 1.0  & 52   \\
3 & 1.2 & 8.60    & 14.5   &  9e-09     & 34.0 & 1.0  & 52   \\
4 & 1.2 & 8.60    & 14.5   &  5e-09     & 27.5 & 1.0  & 52   \\
5 & 1.0 & 8.78    & 18.0   &  2e-11     & 20.0 & 0.2  & --   \\
\enddata
\tablenotetext{1}
{The initial temperature of the white dwarf.
}
\tablenotetext{2}
{The maximum temperature reached by the model during the run.
}
\tablenotetext{3}
{Models for which the effect of boundary layer irradiation is taken into
account $(\Omega_{wd} < 1 )$ are run for a few days only.
}

\end{deluxetable}

\clearpage

{\large{\bf{\center{
Figures Caption }}}} 

Figure 1: The VSNET optical light curve of the 2001 outburst of WZ Sge.
The first 6 observations are marked by arrows. The four different phases
of WZ Sge have been marked for clarity.  

Figure 2: Modeling the heating and cooling of WZ Sge. 
The temperature (in Kelvin) of the
white dwarf is drawn as a function of time (in days) since the start of
the outburst (July 23, 2001)
The solid line represents the compressional 
heating modeling, for 1.2 solar mass white dwarf with an initial temperature 
of 14,500K, accreting at a rate of $9 \times 10^{-9} M_{\sun}/$year 
for 52 days.
The stars, squares and plus signs denote the temperatures listed
in Table 2 $T_a$, $T_b$ and $T_c$ respectively. 
The triangles are the FUSE data points and the circles represent the 
temperature estimates using the flux values (see text). 

Figure 3: Same as Figure 2, but here the solid line represent a model
with a mass accretion rate of $5 \times 10^{-9} M_{\sun}/$year 
and an initial temperature of 15,000K (model 4 in Table 3). 

Figure 4: Same as Figures 2 and 3. The solid line represents
model 2 in Table 3. 

\clearpage

\begin{figure}
\plotone{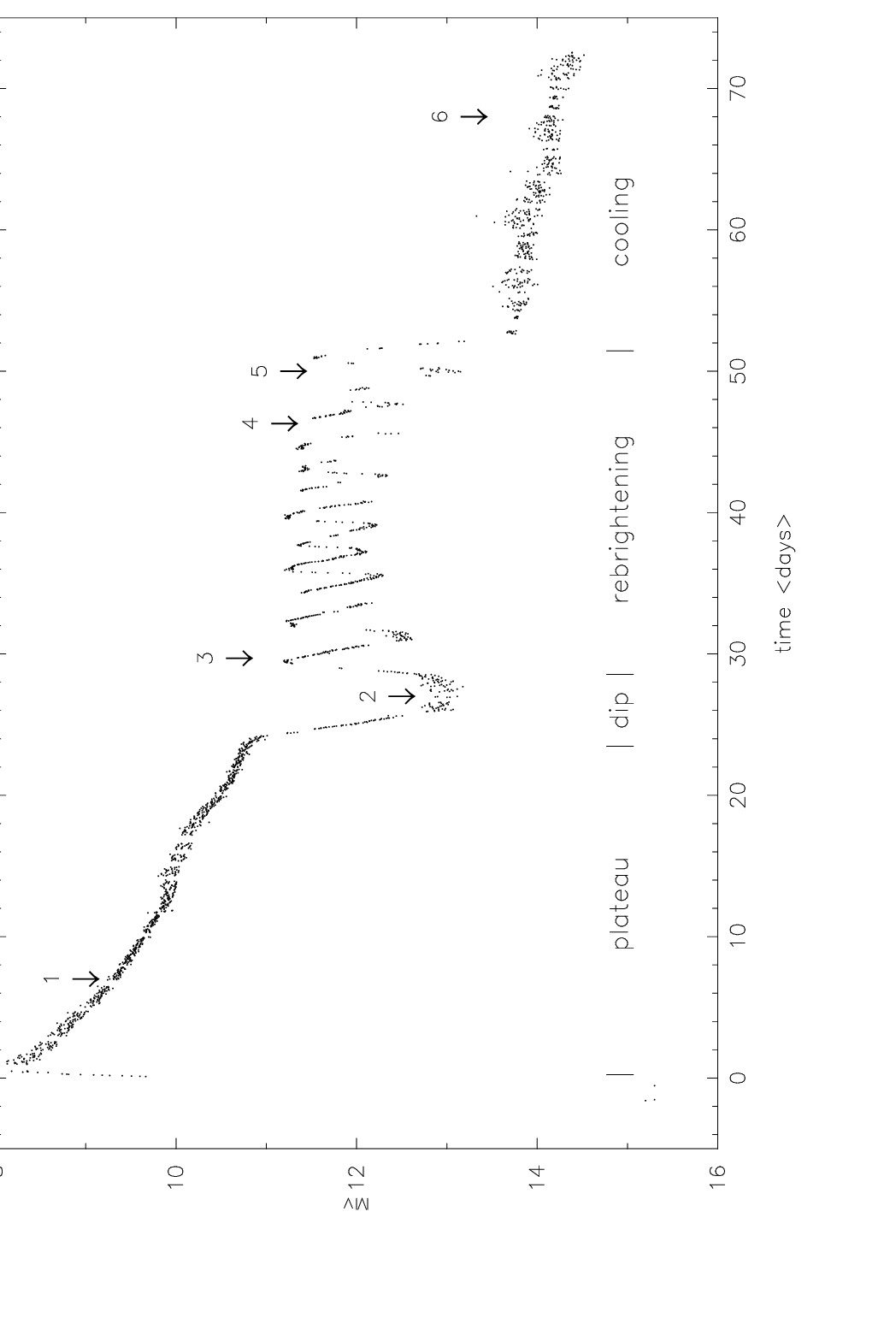}
\figurenum{1}
\end{figure}

\clearpage

\begin{figure}
\plotone{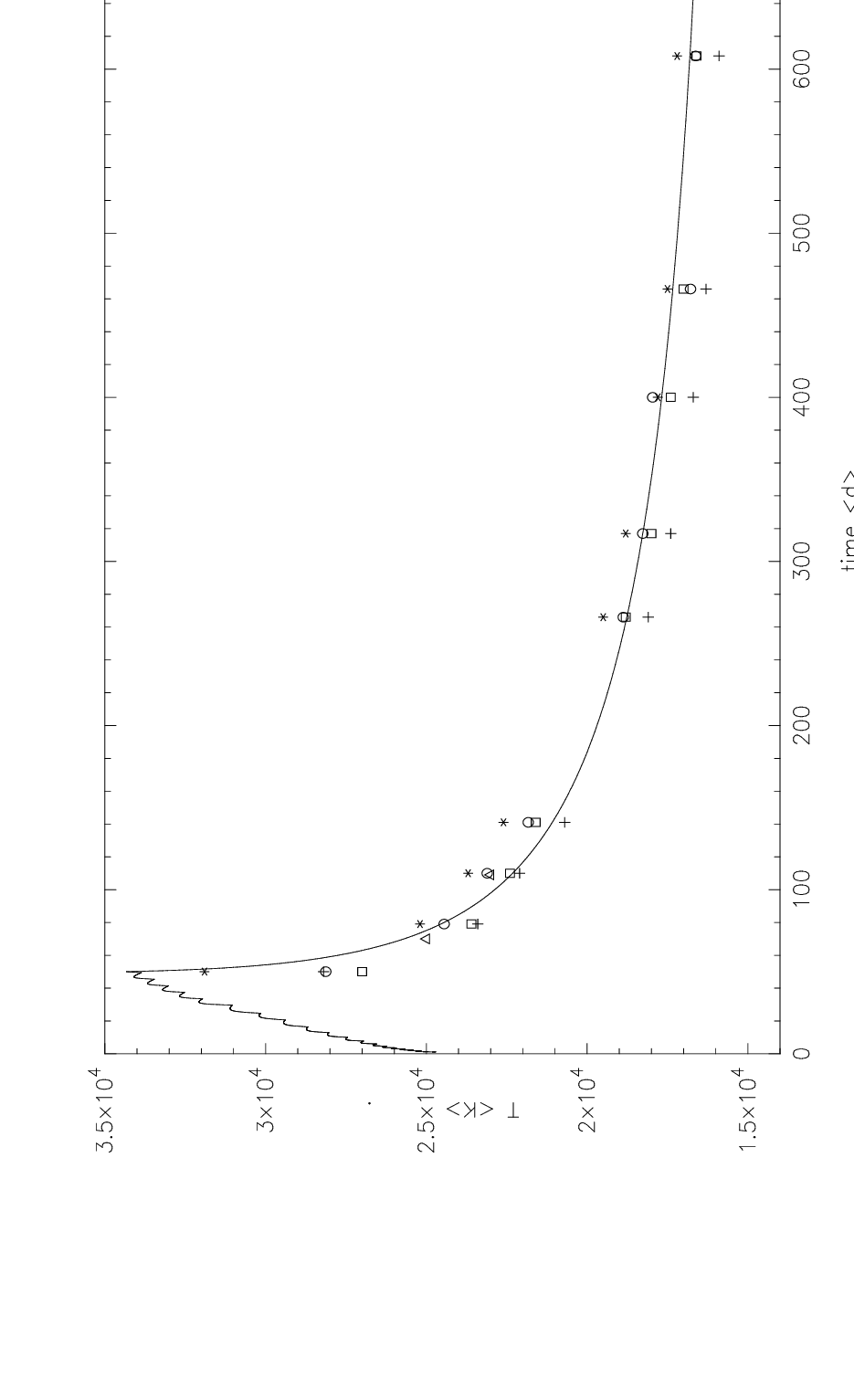}
\figurenum{2}
\end{figure}

\clearpage

\begin{figure}
\plotone{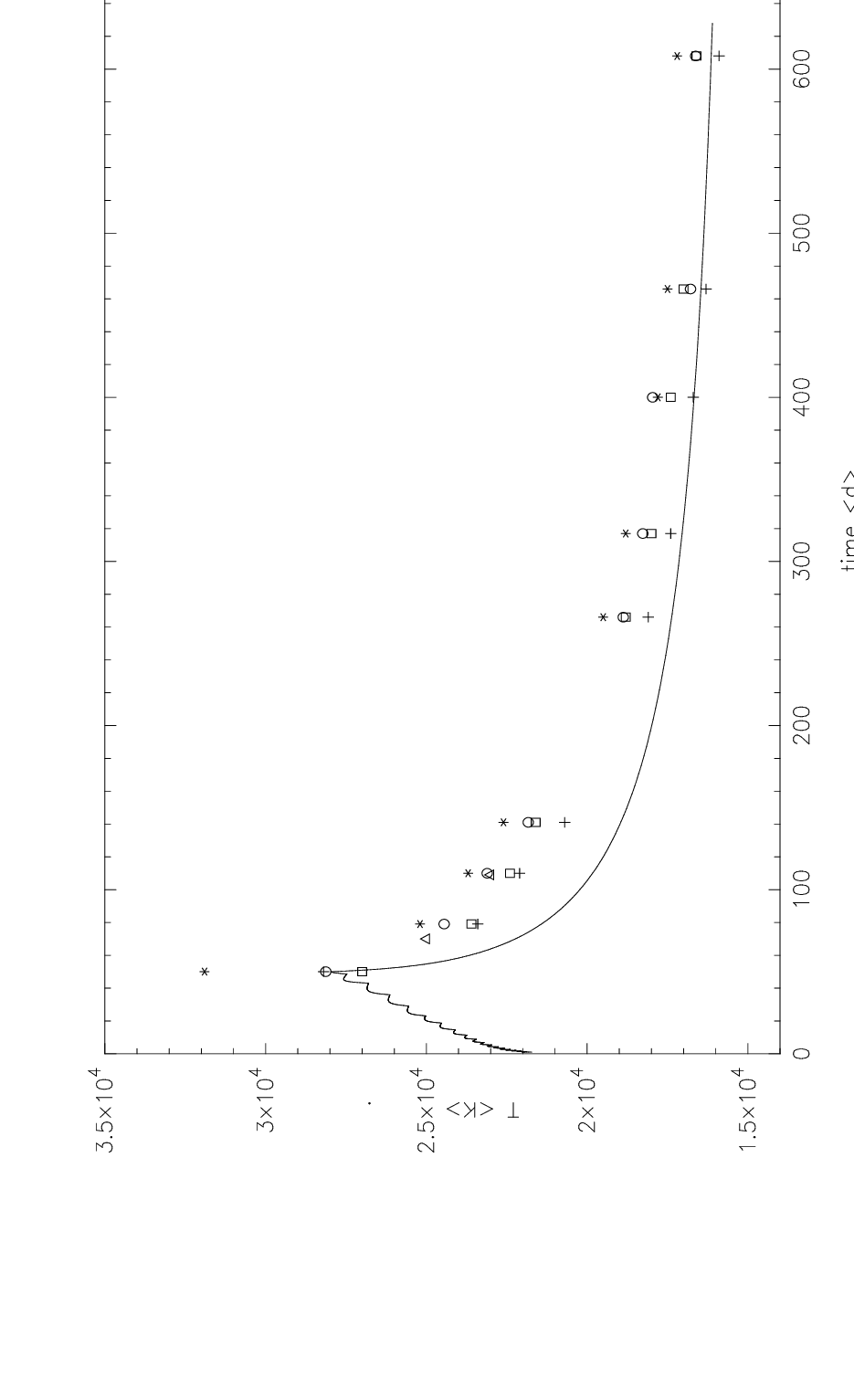}
\figurenum{2}
\end{figure}

\clearpage

\begin{figure}
\plotone{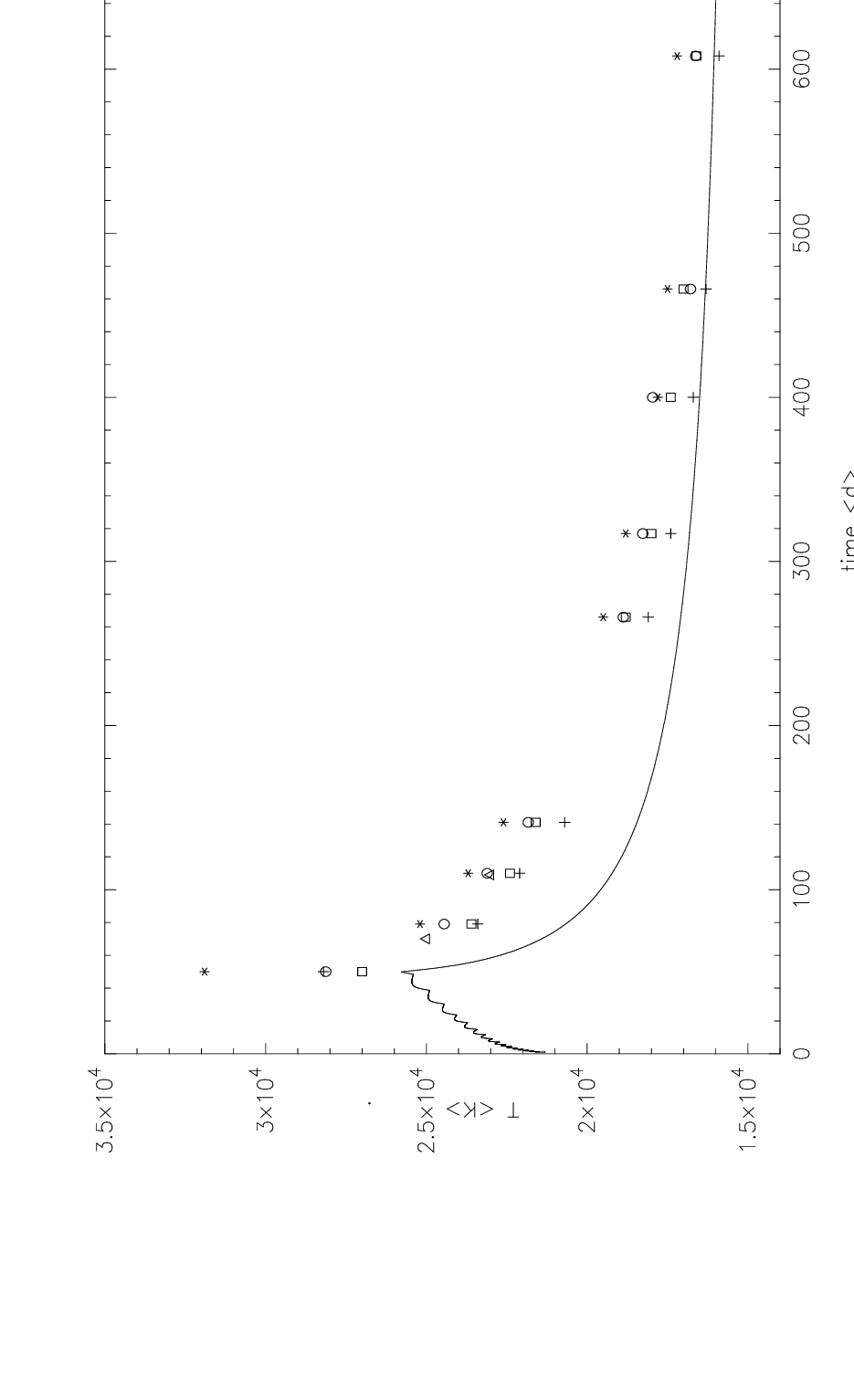}
\figurenum{2}
\end{figure}


\newpage 

\begin{thebibliography}{}

\bibitem[Bailey (1979)]{bai79} 
Bailey, J., 1979, \mnras, 189, 41P 

\bibitem[Cheng et al. (1997)]{che97}
Cheng, F., Sion, E.M., Szkody, P., \& Huang, M. 1997,
ApJL, 484, L149 

\bibitem[Godon \& Sion (2002)]{god02} 
Godon, P., \& Sion, E. M. 2002, \apj, 566, 1084.


\bibitem[Hasenkopf \& Eracleous (2002)]{has02}
Hasenkopf, C.A, \& Eracleous, M., 2002, AAS Meeting 201, 120.03 

\bibitem[Howell et al. (1999)]{how99} 
Howell, S.B., Ciardi, D., Szkody, P., van Paradijs, J., Kuulkers, E.,
Cash, J., Sirk, M., \& Long, K.S. 1999, PASP, 111, 342

\bibitem[Howell et al. (2002)]{how02} 
Howell, S.B., Fried, R., Szkody, P., Sirk, M., 
\& Schmidt, G.2002, PASP, 114, 748

\bibitem[Hubeny (1988)]{hub88} Hubeny, I. 1988, Comput. Phys. Comm., 52, 103

\bibitem[Hubeny et al. (1994)]{hub94} Hubeny, I., Lanz, T.,
\& Jeffrey, S. 1994, Newsletter on Analysis
of Astronomical Spectra (St. Andrews Univ.), 20, 30

\bibitem[Hubeny \& Lanz (1995)]{hub95} Hubeny, I., \& Lanz, T. 1995, ApJ,
439, 875

\bibitem[Ishioka et al. (2001)]{ish01}
Ishioka, R. et al. 2001, IAU Circ., 7669, 1

\bibitem[Klu\'zniak (1987)]{klu87} 
Klu\'zniak 1987, PhD Thesis, Standford University

\bibitem[Knigge et al. (2002)]{kni02} 
Knigge, C., Hynes, R.I., Steeghs, D., Long, K.S., Araujo-Betancor, S.,
\& Marsh, T.R., 2002, ApJ Letters, in press.

\bibitem[Long et al. (2003a)]{lon03a} 
Long, K.S., Froning, C.S., G\"ansicke, B., Knigge, C., Sion, E.M., 
\& Szkody, P.  2003, ApJ, submitted.

\bibitem[Long et al. (2003b)]{lon03b} 
Long, K.S., Sion, E.M., G\"ansicke, B., \& Szkody, P. 2003, submitted to ApJ 

\bibitem[Patterson et al. (2002)]{pat02}
Patterson J., et al. 2002, PASP, 114, 721 

\bibitem[Popham (1999)]{pop99}
Popham, R. 1999, \mnras, 308, 979 

\bibitem[Regev \& Shara (1989)]{reg89} 
Regev, O., \& Shara, M. M. 1989, \apj, 340, 1006 

\bibitem[Shaviv \& Starrfield (1987)]{sha87} 
Shaviv, G., \& Starrfield, S. 1987, \apj, 321, L51 

\bibitem[Sion (1995)]{sio95} 
Sion, E. M. 1995, \apj, 438, 876 

\bibitem[Sion et al. (2003a)]{sio03} 
Sion, E.M., G\"ansicke, B.T., Long, K.S., Szkody, P., Cheng, F., 
Howell, S.B., Godon, P., Knigge, C., Marsh, T., Sparks, W.M., 
\& Starrfield, S., 2003, ApJ, in press.  

\bibitem[Slevinsky et al. (1999)]{sle99}
Slevinsky, R.J., Stys, D., West, S., Sion, E.M., \&
Cheng, R.H. 1999, PASP, 111, 1292 

\bibitem[Sparks et al. (1993)]{spa93}
Sparks, W.M., Sion, E.M., Starrfield, S., Austin, S. 1993, in 
The Physics of Cataclysmic Variables and Related Objects
(eds. Regev and Shaviv)  

\bibitem[Wade \& Hubeny (1998)]{wad98} 
Wade, R., \& Hubeny, I. 1998, ApJ, 509, 350

\end{thebibliography}
\end{document}